\documentclass[twocolumn,aps,superscriptaddress,prb,longbibliography]{revtex4-1}
\usepackage{graphicx}
\usepackage{amssymb}
\usepackage{amsmath}
\usepackage{bm}
\usepackage{color}

\def\Journal #1,#2,#3,#4#5#6#7{#1 {\bf #2}, #3 (#4#5#6#7)}

\def\Vec{\mathbf}

\def\lsim{\, \lower -0.3ex \hbox{$<$} \kern -0.75em \lower 0.7ex \hbox{$\sim$} \,}
\def\gsim{\, \lower -0.3ex \hbox{$>$} \kern -0.75em \lower 0.7ex \hbox{$\sim$} \,}

\usepackage{gensymb}
\newcommand{\iu}{{i\mkern1mu}}

\begin{document}

\title{Perfect one-dimensional chiral states in biased twisted bilayer graphene\\
}

\author{Bonnie Tsim}
\affiliation{National Graphene Institute, University of Manchester, Manchester M13 9PL, UK}
\affiliation{Department of Physics and Astronomy, University of Manchester, Manchester M13 9PL, UK}

\author{Nguyen N. T. Nam}
\affiliation{Mathematics for Advanced Materials Open Innovation Lab (MathAM-OIL), AIST, Sendai 980-8577, Japan}

\author{Mikito Koshino}
%\thanks{koshino@phys.sci.osaka-u.ac.jp}
\affiliation{Department of Physics, Osaka University,  Toyonaka 560-0043, Japan}

\begin{abstract}
We theoretically study the electronic structure of small-angle twisted bilayer graphene
with a large potential asymmetry between the top and bottom layers. 
We show that the emergent topological channels known to appear on the triangular AB-BA domain boundary do not actually form 
a percolating network,
but instead they provide independent, perfect one-dimensional eigenmodes propagating in 
three different directions.
Using the continuum-model Hamiltonian, we demonstrate 
that an applied bias causes two well-defined energy windows 
which contain sparsely distributed one-dimensional eigenmodes.
The origin of these energy windows can be understood using a two-band model
of the intersecting electron and hole bands of single layer graphene.
We also use the tight-binding model to implement the lattice deformations in twisted bilayer graphene, 
and discuss the effect of lattice relaxation on the one-dimensional eigenmodes.
\end{abstract}

\maketitle

\section{Introduction}
Twisted bilayer graphene (TBG) consists of two layers of graphene overlaid on top of each other with a relative twist between their crystallographic axes. 
A moir\'{e} interference pattern which emerges from the overlap of the two mismatched graphene lattices
results in a strong modification of the electronic structure by the superlattice band folding 
\cite{lopes2007graphene,mele2010commensuration,trambly2010localization,shallcross2010electronic,
morell2010flat,bistritzer2011moirepnas,kindermann2011local,PhysRevB.86.155449,moon2012energy,de2012numerical}.
The system has been shown to exhibit many interesting physical phenomena and, since the realisation of superconductivity and correlated insulating states in magic-angle TBG \cite{cao2018unconventional,cao2018mott,yankowitz2019tuning}, there has been a huge surge of theoretical and experimental research in this field. 

\begin{figure}
\center{
\includegraphics[width=1.\hsize]{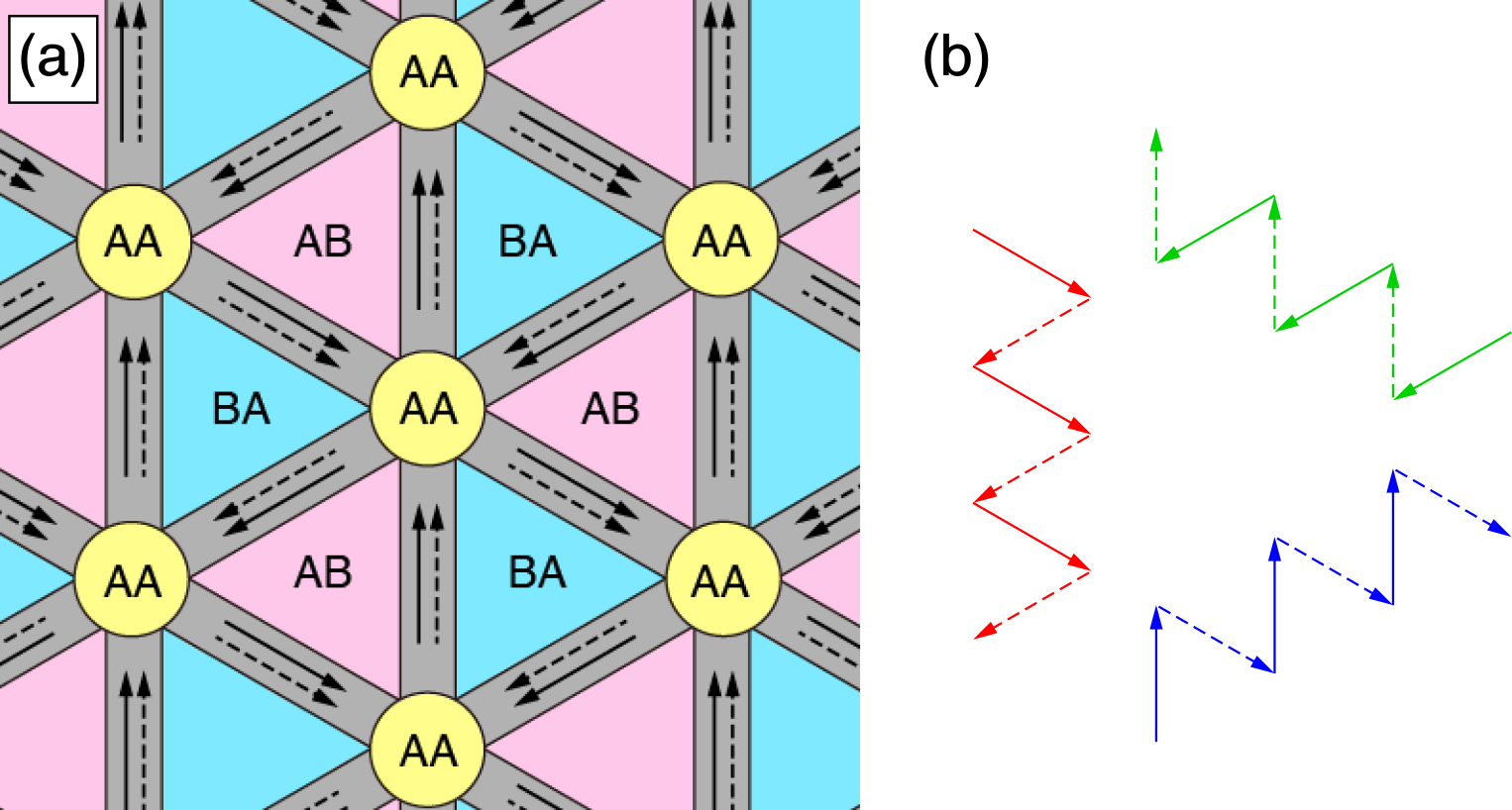}
}
\caption{(a) AB and BA domain structure and topological boundary channels  in the biased TBG.
Here solid and dashed arrows represent independent traveling modes for $K$ valley (mode 1 and 2).
(b) Independent 1D eigenmodes in three directions.} 
\label{fig_schem_domain}
\end{figure}

In this paper, we theoretically study the electronic structure of small-angle TBG 
with a large interlayer bias (i.e., potential asymmetry between the top and bottom layers), 
and demonstrate the formation of perfect one-dimensional (1D) states 
within well-defined energy windows on either side of zero energy.
The effect of the interlayer bias on TBG was investigated in the previous theoretical works
\cite{xian2011effects,san2013helical,moon2014optical,ramires2018electrically,efimkin2018helical,fleischmann2019perfect, walet2019emergence,hou2019current},
and it was found that a large enough bias gives rise to a network of topological channels on the  
domain boundaries between AB and BA stacking regions \cite{san2013helical,ramires2018electrically,efimkin2018helical,fleischmann2019perfect,walet2019emergence,hou2019current}.
There the electronic states at AB and BA regions are locally gapped out 
by the interlayer bias \cite{mccann2006asymmetry}, and 
two topological modes per spin and per valley necessarily appear on each AB-BA boundary 
\cite{vaezi2013topological,zhang2013valley,pelc2015topologically,ju2015topological,yin2016direct,lee2016zero,li2016gate},
reflecting that the two regions have different quantized values of single-valley Hall conductivity,
$\pm e^2/h$. \cite{koshino2008electron}.
In TBG, the AB and BA regions appear periodically in a hexagonal pattern \cite{brown2012twinning,lin2013ac,alden2013strain}
such that the boundary channels form a triangular grid as illustrated in Fig.\ \ref{fig_schem_domain}(a).
Recently, the network of the topological channels in TBG was experimentally probed 
by transport measurements \cite{rickhaus2018transport,yoo2019atomic,giant2019oscillations}
and also by scanning tunneling spectroscopy \cite{huang2018topologically}.

One may expect that the electronic transport in the topological channels of TBG could be described by 
a percolation model \cite{chalker1988percolation} on a triangular network.
However, here we show that the topological modes do not form a two-dimensional network, but they are actually independent 1D eigenmodes composed of a serial connection of topological channel sections
as shown in Fig.\ \ref{fig_schem_domain}(b).
The modes along different directions are never hybridized, and therefore
all these states serve as independent perfect 1D channels over the whole TBG.
The result is consistent with a recent work which predicts the perfect nesting of the Fermi surface
in the biased TBG \cite{fleischmann2019perfect}.

In the following, we calculate the electronic band structure of the TBG 
using the continuum-model Hamiltonian 
and present the band structures for various twist angles and electric field dependencies. 
The energy band structures show that an applied bias causes two well-defined energy windows 
which contain sparsely distributed perfect 1D eigenmodes,
separated by a cluster of nearly flat bands around the charge neutrality point. 
We also use the tight-binding model to implement arbitrary lattice deformations in TBG, 
and discuss the effect of lattice relaxation on the 1D eigenmodes.
Lastly, we explain the origin of these energy windows 
by a perturbational approach from the small interlayer coupling limit,
and also by a two-band model consisting of the intersecting electron and hole bands of single layer graphene.
The tunability of the TBG energy dispersion in a perpendicular electric field means there is the potential to explore the parameter space where these 1D eigenmodes can be found in its experimental realization.

This paper is organized as follows: In Sec. II, we introduce the continuum-model Hamiltonian and describe the formation of perfect 1D eigenmodes for varying angles and biases. In Sec. III, we consider the effect of lattice relaxation on the 1D eigenmodes. Lastly, we explain the origin of 1D eigenmodes in Sec. IV, and present a brief conclusion in Sec. V.

%The purpose of this work is to investigate the origin of the 1D states of TBG under large biases. Previous works have shown that in marginally twisted bilayer graphene under small biases, different stacking arrangements within the moir\'{e} interference pattern become separated by domain walls where electrons are localised. This results in a triangular network of states that can be observed at low energies [?].

 \begin{figure*}
        \center{
        \includegraphics[width=18cm]{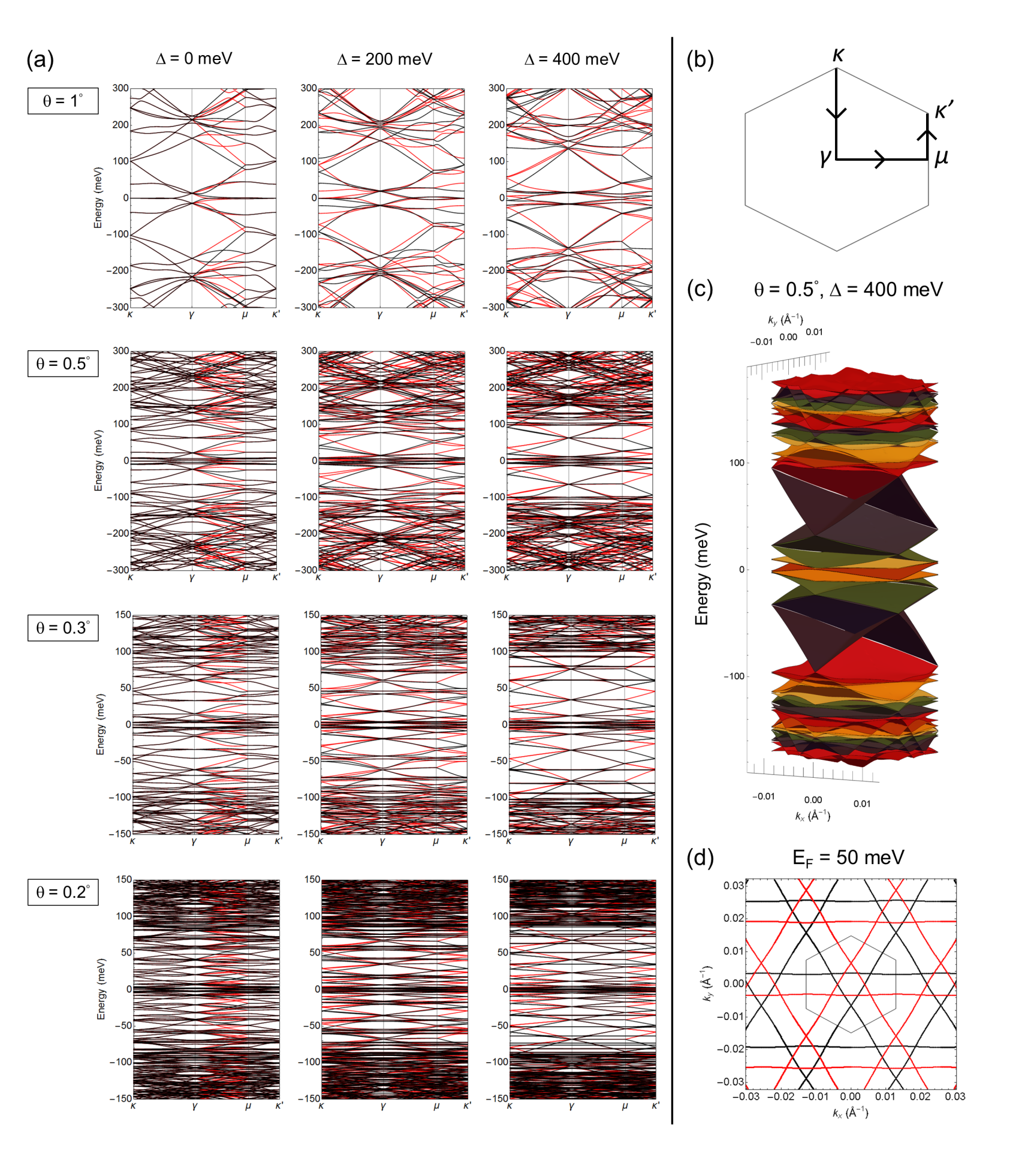}
        }
        \caption{(a) Band structure of the twisted bilayer at various twist angles and varying $\Delta$, calculated using the continuum model. (b) The moir\'e Brillouin zone.
(c) A three dimensional plot of $K$-valley bands, and (d) the contour plot at $E_F= 50$ meV,
calculated for  $\theta = 0.5 \degree$ and $\Delta = 400$ meV.
The black and red lines represent $K$ and $K'$ valleys, respectively.} 
        \label{fig_band}
      \end{figure*}
     
\section{Continuum Model}

We calculate the electronic band structure of the twisted bilayer graphene 
using the continuum model \cite{lopes2007graphene,bistritzer2011moirepnas,kindermann2011local,PhysRevB.86.155449,moon2013opticalabsorption,koshino2015interlayer,koshino2015electronic}.
For a small twist angle, the Hamiltonian is given by
\begin{eqnarray}
	H_{\text{TBG}} = 
	\begin{pmatrix}
		H_1 & U^\dagger \\
		U & H_2
	\end{pmatrix},
	\label{eq_H_eff}
\end{eqnarray}
where
\begin{align}
& H_1 = \begin{pmatrix}
\frac{\Delta}{2} & -\upsilon\pi^{\dagger} \\
-\upsilon \pi & \frac{\Delta}{2}
\end{pmatrix},
\quad
H_2 = \begin{pmatrix}
- \frac{\Delta}{2} & -\upsilon\pi^{\dagger} \\
-\upsilon \pi & - \frac{\Delta}{2}
\end{pmatrix} 
\label{eq_intra}
\end{align}
and 
\begin{align}
 U &= 
 u\displaystyle\sum_{j=0,1,2} e^{\iu \Delta \mathbf{K}_j \mathbf{\cdot r}} 
 \begin{pmatrix}
1 & e^{-\iu\frac{2\pi}{3}j} \\
e^{\iu\frac{2\pi}{3}j} & 1
\end{pmatrix}. 
\label{eq_interlayer_matrix}
\end{align}
The Hamiltonian Eq.\ (\ref{eq_H_eff}) is equivalent to the continuum-model Hamiltonian derived 
in \cite{lopes2007graphene,bistritzer2011moirepnas, moon2013opticalabsorption} up to a gauge transformation.\cite{tarnopolsky2019origin,liu2019pseudo} 
The on-diagonal blocks describe the graphene layers 1 and 2 where $\pi= \hbar(\xi k_x + \text{i} k_y) $, 
and the valley index $\xi=\pm 1$ is used to distinguish between $K$ and $K'$ valleys. 
The parameter $\upsilon$ is the band velocity of monolayer graphene where $\hbar\upsilon/a = 2.1354$~eV (the lattice constant of graphene is given by $a=2.46$~\text{\AA}) \cite{moon2013opticalabsorption, koshino2018maximally}, and $\Delta$ represents the electrostatic energy shift induced by the perpendicular electric field. The off-diagonal blocks describe the moir\'e interlayer coupling between the two twisted layers, where the interlayer coupling strength is given by $u=0.103$~eV. The vectors $\Delta \mathbf{K}_j\,(j=0,1,2)$ account for the shift between the original Brillouin zone corners of the two layers, and are given by
\begin{equation}
\Delta \mathbf{K}_{j} = \frac{4 \pi\theta }{3a} \Big[ -\sin\left(\frac{2\pi j}{3}\right), \cos\left(\frac{2\pi j}{3}\right) \Big],
\end{equation}
where  $\theta$ is the twist angle between the two layers in radians.

We calculate the energy spectrum for the $K$ and $K'$ valleys independently as intervalley coupling is negligible at small twist angles. Zone folding is used to bring the states in each valley with momenta connected by the moir\'{e} reciprocal lattice vectors,
$\mathbf{G}_1 = \Delta\mathbf{K}_1 - \Delta\mathbf{K}_0$ and  $\mathbf{G}_2 = \Delta\mathbf{K}_2 - \Delta\mathbf{K}_0$.
The basis of $k$-states of layer 1 and 2 can be taken as
\begin{align}
&\Vec{k} ^{(1)}_{m_1,m_2} = \Vec{k} + \Delta \mathbf{K}_0 + m_1 \mathbf{G}_1 + m_2 \mathbf{G}_2 \nonumber\\ 
&\Vec{k} ^{(2)}_{m_1,m_2} = \Vec{k} - \Delta \mathbf{K}_0 + m_1 \mathbf{G}_1 + m_2 \mathbf{G}_2,
\label{eq_k_vectors}
\end{align} 
respectively, where 
$\Vec{k}$ is the wavenumber in the first moir\'{e} Brillouin zone (mBZ) spanned by  $\mathbf{G}_1$ and  $\mathbf{G}_2$,  and $m_1$ and $m_2$ are integers. The size of the basis is chosen such that when the Hamiltonian is numerically diagonalized, the energy bands converge up to a cut-off energy.

Figure \ref{fig_band}(a) presents the electric field dependence of the TBG band structure for various twist angles, $\theta = 1\degree, 0.5\degree, 0.3\degree$ and $0.2\degree$. 
The band structures include energy bands from both the $K$ (black) and $K'$ (red) valleys 
and is shown for the path $\kappa \rightarrow \gamma \rightarrow \mu \rightarrow\kappa'$ in the mBZ illustrated in Fig.\ \ref{fig_band}(b).
The original Dirac point of layer 1 is placed at the corner of mBZ at $\kappa'$, and the original Dirac point of layer 2
is placed at $\kappa$. 
In increasing $\Delta$, we see that the energy bands are gradually shifted toward zero energy, 
forming a cluster of nearly flat bands
around the charge neutrality point. At the same time, two well-defined energy windows, 
where energy bands are only sparsely distributed, 
are formed above and below the zero-energy band cluster.
The size of the energy windows are not strongly affected by the size of $\Delta$ which can be seen for $\theta = 0.2\degree$ in increasing $\Delta$. 
%For a fixed $\Delta$, the size of the energy window has a weak dependence on the twist angle. For instance, when $\Delta = 400$~meV, the size of the energy window decreases gradually from approximately $150$ to $90$~meV as the twist angle decreases from $\theta = 1\degree$ to  $0.2\degree$.

Most interestingly, is the formation of 1D propagating modes inside the energy windows,
which connect the zero-energy band cluster to the bulk bands outside of the energy windows.
Figure \ref{fig_band}(c) shows a three dimensional plot of the bands from $K$ valley 
calculated for  $\theta = 0.5 \degree$ and $\Delta = 400$ meV,
and Fig. \ref{fig_band}(d) is the Fermi surface of the same system at $E_F= 50$ meV where
black and red lines represent $K$ and $K'$, respectively.
We see that the band dispersion of $K$ is actually composed of three intersecting planes,
with band velocities parallel to $(0,-1)$, $(\sqrt{3}/2,1/2)$ and $(-\sqrt{3}/2,1/2)$ directions.
The different planes are not hybridized with each other, giving nearly straight Fermi lines at the fixed energy.
Such straight Fermi surfaces were also reported in the recent paper \cite{fleischmann2019perfect}.
In the largest bias $\Delta=400$ meV, we notice 
some flat levels appear in the upper part of the energy window 
independently from the dispersive 1D states,
(e.g., three horizontal lines in 50 meV $< |E| <$ 100 meV for $\theta = 0.3^\circ$),
which can be interpreted as pseudo-Landau levels of the fictitious gauge field \cite{ramires2018electrically}.

%%%%
 \begin{figure*}
        \center{
        \includegraphics[width=0.7\hsize]{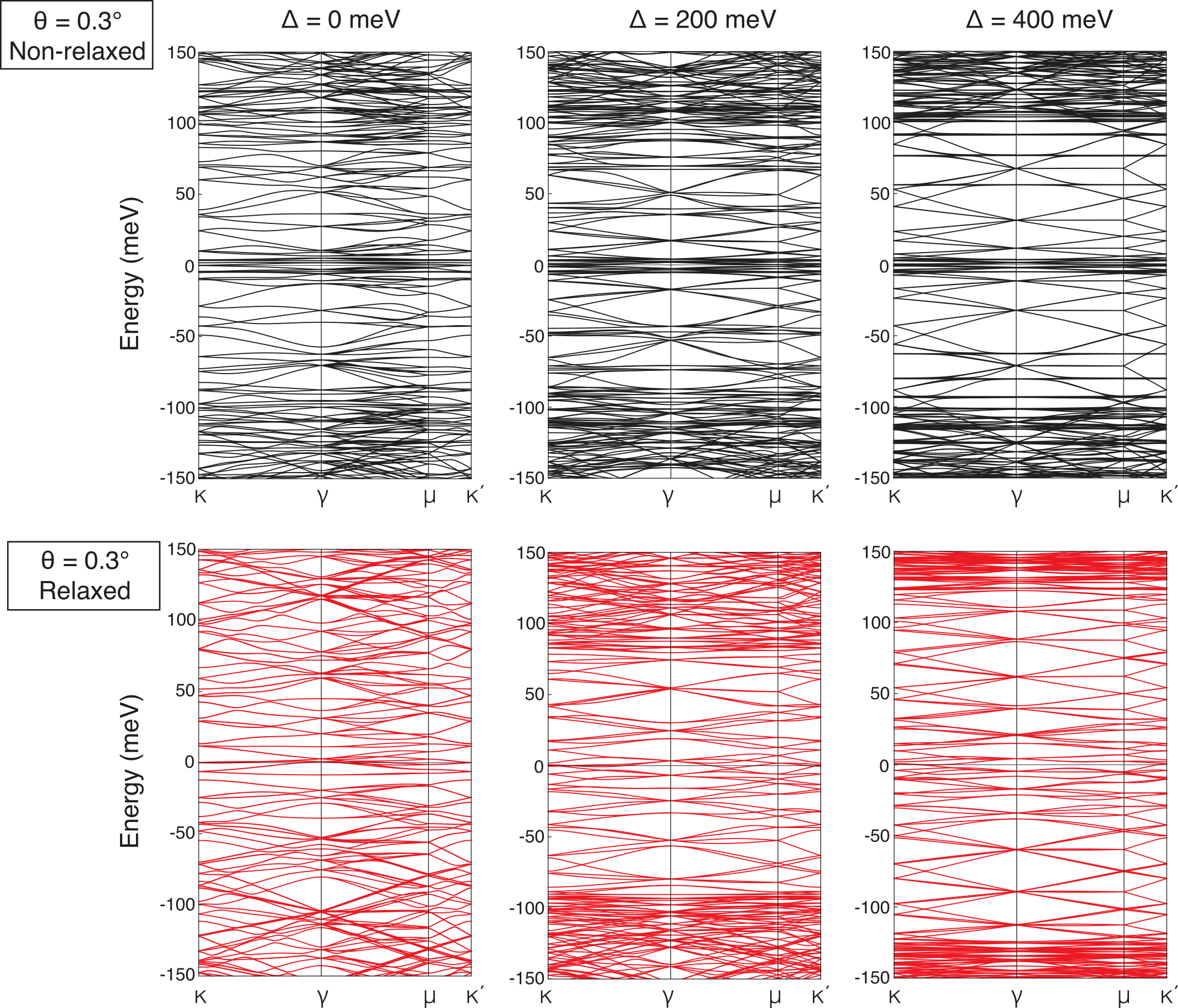}
        }
\caption{Band structure of non-relaxed (upper panels) and relaxed (lower panels) TBGs
in $\theta =0.3^\circ$ and different $\Delta$'s and calculated using the tight-binding model.}
        \label{fig_band_relax}
      \end{figure*}

\begin{figure}
        \center{
        \includegraphics[width=0.9\hsize]{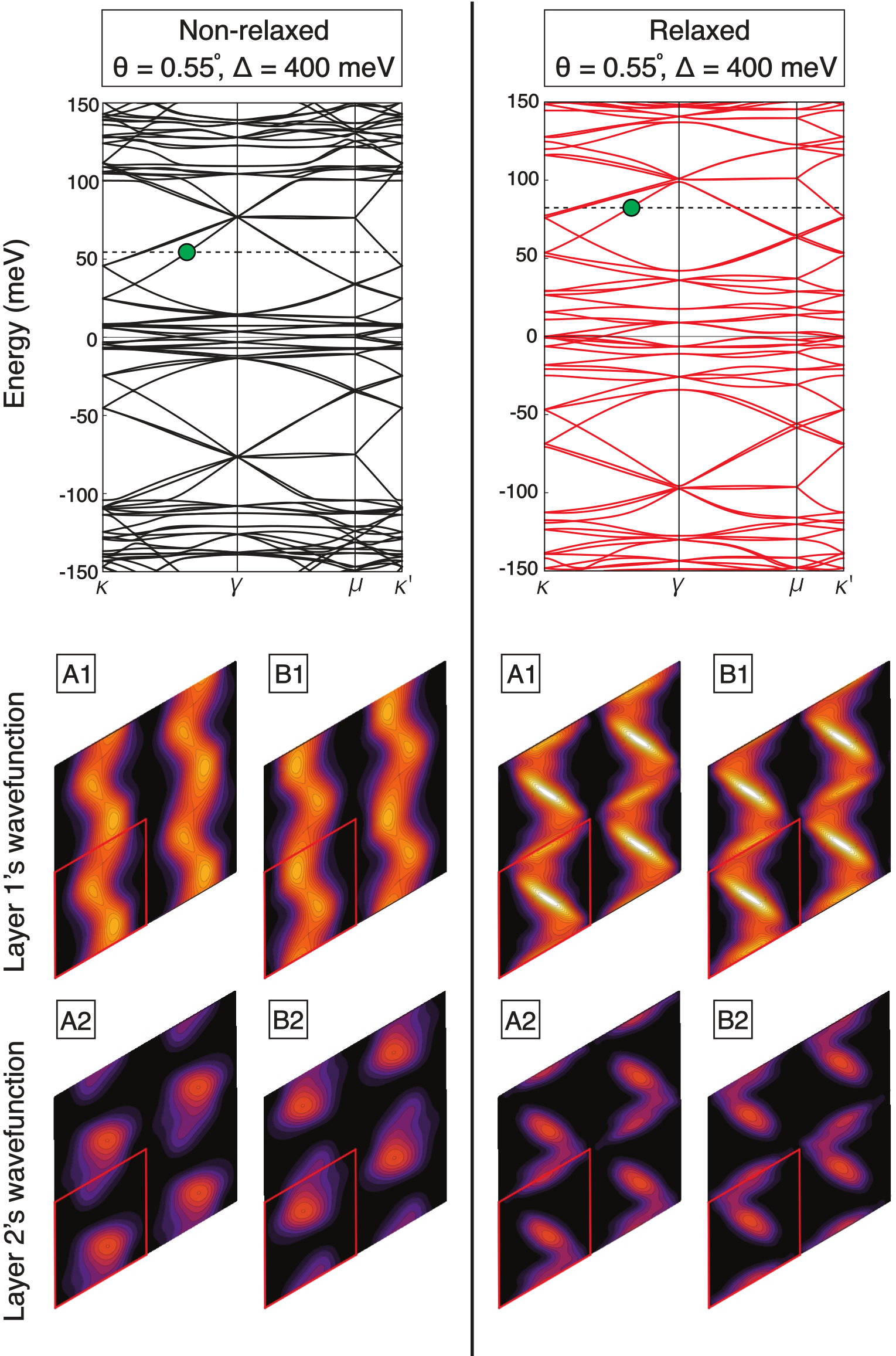}
        }
        \caption{
        Energy spectrum and the wave function of a typical state at $K$ valley in the energy window
        (indicated in the spectrum) calculated for the non-relaxed TBG and the relaxed TBG 
        of $\theta = 0.55^\circ$ and $\Delta = 400$ meV. The plot of the wave function represents squared amplitude on sublattice
        $A_1$, $B_1$, $A_2$ and $B_2$ separately, where a red rhombus indicates a single moir\'{e} unit cell.}
        \label{fig_wave}
      \end{figure}

 %%%
 
\section{Effect of lattice relaxation}

The real TBG is not a simple stack of rigid graphene layers as assumed in the previous section, but
it has a spontaneous lattice relaxation and resulting AB/BA domain formation
 \cite{popov2011commensurate, brown2012twinning,lin2013ac,alden2013strain,uchida2014atomic,
van2015relaxation,dai2016twisted, jain2016structure,nam2017lattice,carr2018relaxation, lin2018shear,yoo2019atomic,guinea2019continuum}.
Such a structural deformation modifies the electronic band structure \cite{nam2017lattice, lin2018shear, koshino2018maximally, yoo2019atomic,guinea2019continuum,crucial2019lucignano,fleischmann2019perfect, walet2019emergence}.
Here we calculate the energy band structures in the presence of the lattice strain
using the tight-binding method \cite{nam2017lattice}. 
The Hamiltonian is given by
\begin{equation}
H = -\sum_{i,j} t(\textbf{R}_i -\textbf{R}_j) |\textbf{R}_i \rangle \langle \textbf{R}_j| + \text{h.c.}
\end{equation}
where $\textbf{R}_i$ is the atomic coordinate, $ |\textbf{R}_i \rangle $ is the wave function at site $i$, 
and $t(\textbf{R}_i -\textbf{R}_j)$ is the transfer integral between atom $i$ and $j$. 
We adopt the Slater-Koster type formula for the transfer integral
 \cite{slater1954simplified},
\begin{equation}
-t (\textbf{d}) = V_{pp\pi} (d) \left[1 - \left(\frac{\textbf{d} \cdot \textbf{e}_z}{d}\right)^2 \right] + 
 V_{pp\sigma} (d)  \left(\frac{\textbf{d} \cdot \textbf{e}_z}{d}\right)^2 ,
 \label{eq_t}
\end{equation}
\begin{align}
& V_{pp\pi} (d) = V_{pp\pi}^0  \exp \left(- \frac{d-a_0}{r_0} \right), \\
& V_{pp\sigma} (d)  =V_{pp\sigma}^0  \exp \left(- \frac{d-d_0}{r_0} \right),
\end{align}
where $\textbf{d} = \textbf{R}_i - \textbf{R}_j$ is the distance between two atoms and $\textbf{e}_z$ is the unit vector on $z$ axis. 
$V_{pp\pi}^0 \approx -2.7$ eV is the transfer integral between nearest-neighbor atoms of monolayer graphene which 
are located at distance $a_0 = a/\sqrt{3} \approx 0.142$ nm, $V_{pp\sigma}^0 \approx 0.48 $ eV is the transfer integral 
between two nearest-vertically aligned atoms and $d_0 \approx 0.334$nm is the interlayer spacing. 
The decay length $r_0$ of transfer integral is chosen as $0.184 a$
so that the next nearest intralayer coupling becomes $0.1 V_{pp\pi}^0$.
At $d > \sqrt{3}a$, the transfer integral is very small and negligible.
The optimized atomic positions are obtained by the method introduced in the previous work \cite{nam2017lattice}.
Using this, we construct the tight-binding Hamiltonian of the relaxed TBG and calculate the energy bands.

Figure \ref{fig_band_relax} compares the electronic band structure
of non-relaxed (upper panels) and relaxed (lower panels) TBGs
in $\theta =0.3^\circ$ and different $\Delta$'s.
In the tight-binding model, the valleys are not distinguished.
 We see that the energy bands of the non-relaxed calculation
 quantitatively agree with those in the continuum method in Fig.\ \ref{fig_band}.
 In the presence of the relaxation, we confirm that the qualitative feature remains the same: 
we still see the energy windows and the perfect 1D eigenmodes.
The major difference from the non-relaxed state is that the central pseudo-Landau levels  mentioned in the previous section
are completely hybridized with 1D eigenmodes,
and become a part of the dispersive bands.
Also we notice that the bands in the zero-energy cluster become less flat
and a bit more dispersive.

Figure \ref{fig_wave} shows typical wave functions in the energy window
in the non-relaxed TBG and the relaxed TBG of $\theta = 0.55^\circ$ and $\Delta = 400$ meV.
Here we chose the corresponding states in non-relaxed and relaxed cases,
which are connected by a continuous increase of the relaxation.
The state is chosen from a 1D band of the electron side
with the velocity along $k_y$ axis ($\kappa-\gamma$ direction in this figure).
In each case, we observe that the wave function takes a 1D form extending along $y$
direction, while it is disconnected in the perpendicular direction.
The states in the different Fermi surface branches at the same energy
are obtained by $\pm 120^\circ$ rotation of this figure.
We confirm that the local current density is along $-y$ direction,
in accordance with the negative band velocity in the $k_y$ direction.
The wave amplitude is mainly concentrated on the layer 1, while it is concentrated on layer 2 in the hole side states.
%The different states on the same Fermi-surface branch has almost the same distribution
%of the wave amplitude, while they only differ in the relative phase factor between neighboring channels in the real space.

In the presence of the lattice relaxation, we see that the wave function becomes more localized
on the AB-BA boundary. This is natural because the AB and BA regions
(where the energy band is gapped out) significantly expand under the lattice relaxation, 
and the wave amplitude must be confined to the narrow boundary region.\cite{fleischmann2019perfect, walet2019emergence}
The similar, zigzag-shaped wave function was also reported in a recent work \cite{fleischmann2019perfect}.
Interestingly, we see that the relaxed TBG's wave function in Fig.\ \ref{fig_wave} 
has its amplitude not on the boundary along $y$,
but only on the boundary in the two other directions along $(\sqrt{3}/2, -1/2)$ and $(-\sqrt{3}/2, -1/2)$.
With closer inspection, we also find that
it has different structures between the boundaries along $(\sqrt{3}/2, -1/2)$ and $(-\sqrt{3}/2, -1/2)$,
although the atomic structures are completely symmetric.
Recalling that each single boundary has two different traveling modes (denoted as mode 1 and 2) 
as mentioned, this result suggests that, at every vertex of the triangular grid (AA region), 
mode 1 is always scattered to mode 2 in $-120^\circ$ direction,
while mode 2 is always scattered to mode 1 in $+120^\circ$ direction.
As a result, we have three independent, zigzag traveling modes as illustrated in Fig.\ \ref{fig_schem_domain}(b).

%%%%      
\section{Origin of the perfect 1D eigenmodes}

\begin{figure*}
        \center{
        \includegraphics[width=15cm]{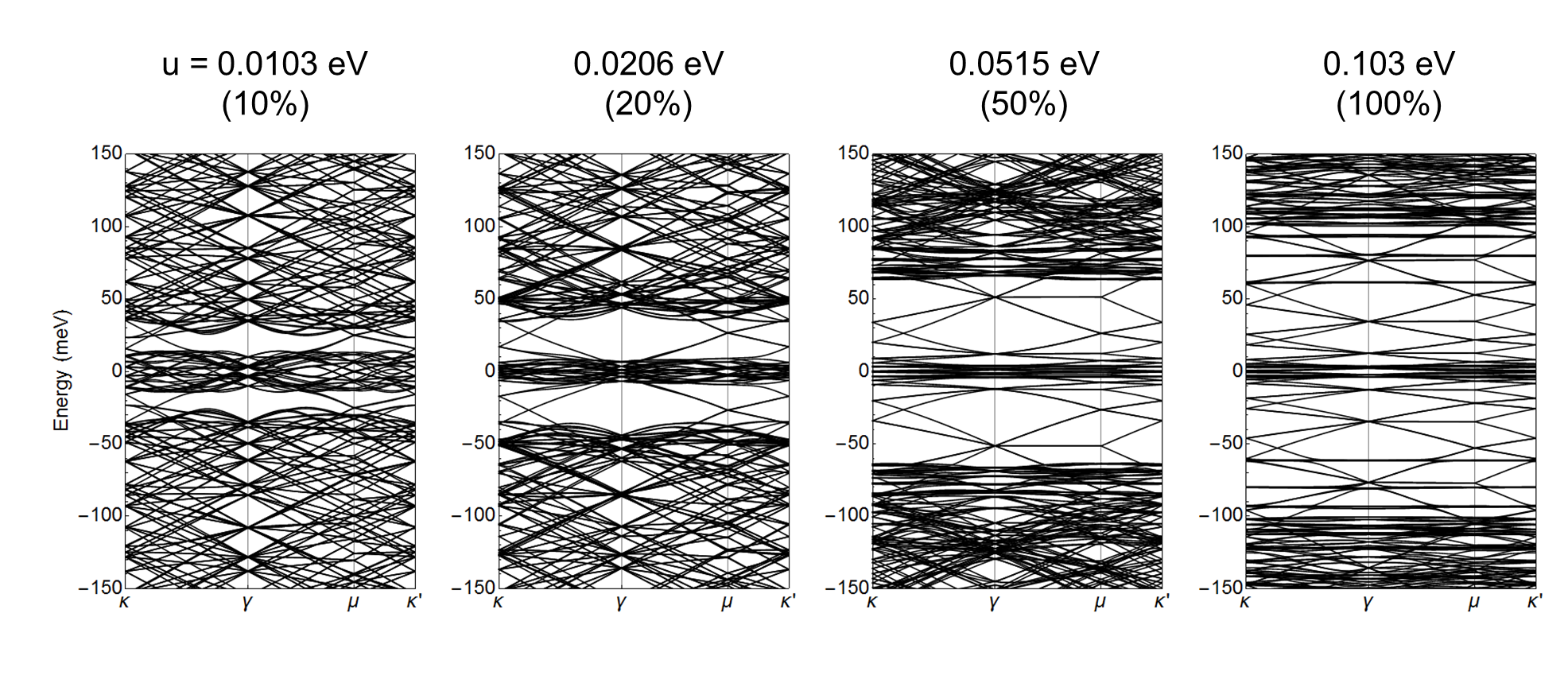}
        }
        \caption{Band structure of the continuum model
for the TBG with $\theta =  0.3 \degree$ and $\Delta=400$ meV, 
and the interlayer coupling $u$ from zero to the actual value in TBG.
} 
        \label{fig_band_u}
\end{figure*}

The origin of the energy window and the 1D eigenmodes
can be intuitively understood by a perturbational approach from the small interlayer coupling limit.
Figure \ref{fig_band_u} shows the band structure of the continuum model in Eq.\ (\ref{eq_H_eff}),
with $\theta =  0.3 \degree$ and $\Delta=400$ meV, and with 
increasing interlayer coupling $u$ from zero to the actual value in TBG.
With small $u$, we see that two gaps open in the electron side and the hole side,
 and they eventually become the window regions in the full $u$ parameter.
We see that the 1D eigenmodes always remain inside the gap, preventing the spectrum from becoming fully gapped. 
The width of the energy window is obviously the order of $u$.
The energy bands between the two gaps are squashed with increasing $u$,
and finally becomes the zero-energy band cluster.

\begin{figure*}
        \center{
        \includegraphics[width=0.8\hsize]{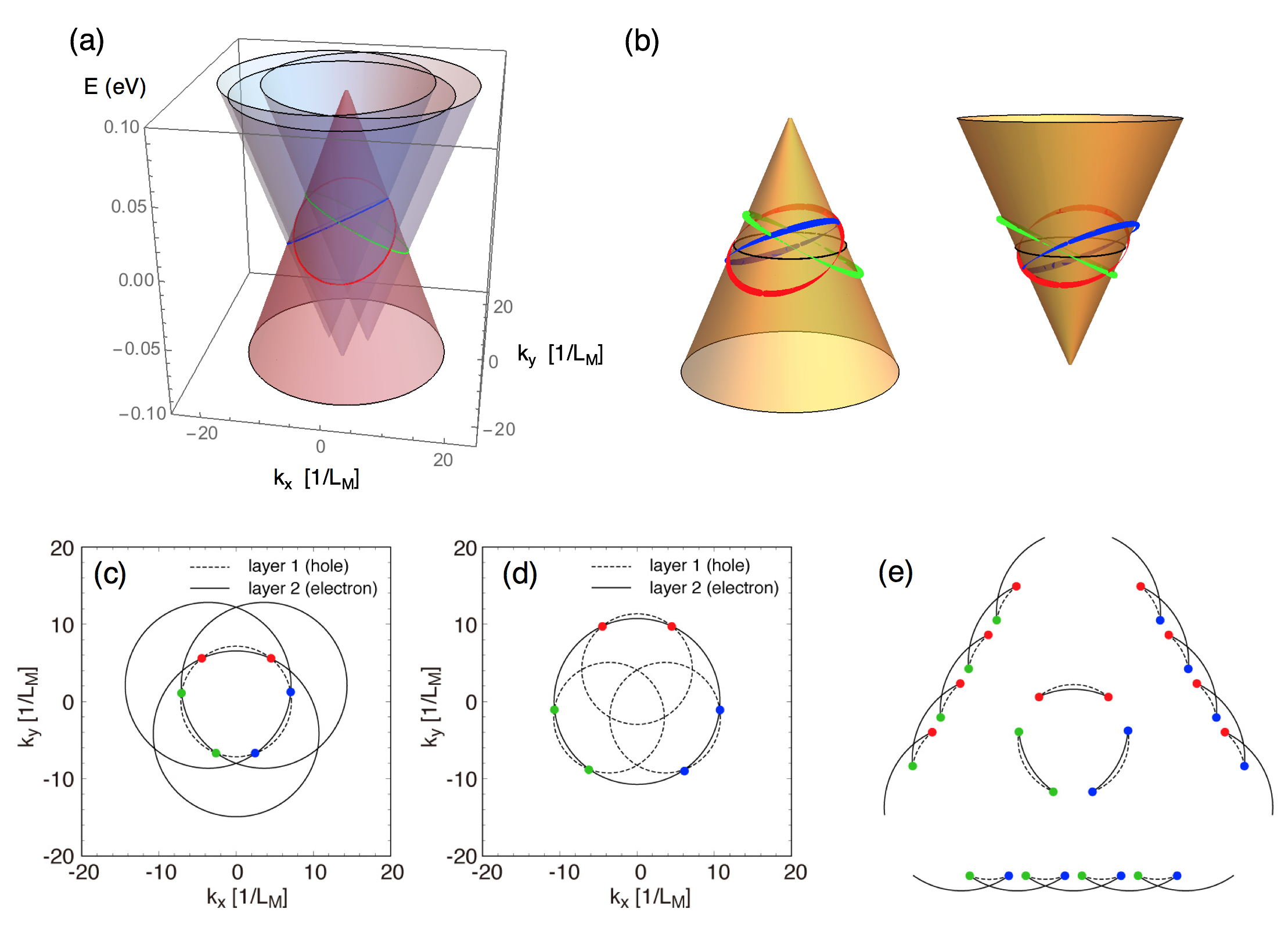}
        }
        \caption{(a) Crossing lines between the Dirac cones,
where three circles (red, blue and green) correspond to $j=0,1,2$.
Here we take $\theta =  0.3 \degree$ and $\Delta=200$ meV.
(b) Amplitude of the interlayer matrix element on the crossing lines on the Dirac cones of layer 1 and 2.
(c), (d) Relative positions of the Fermi surfaces 
of layer 1 (dashed) and layer 2 (solid), in absence of $u$ and at $E_F=20$ meV. (e) The hybridized Fermi surfaces after the infinitesimal anti-crossing.}

        \label{fig_FS}
\end{figure*}

The opening of the two gaps in small $u$ can be explained by considering the following two-band model.
In a large $\Delta$, the low energy region is dominated by the hole band of graphene layer 1 and the electron band of layer 2. 
While considering the interlayer coupling $U$, we can imagine that
the two opposite conical bands of single layer graphenes are 
crossing with each other with a relative momentum shift of $\Delta \Vec{K}_j (j=0,1,2)$,
and the band anti-crossing occurs at the cross section.
Figure \ref{fig_FS}(a) illustrates the actual crossing lines between the Dirac cones
in the case of $\theta =  0.3 \degree$ and $\Delta=200$ meV,
where three circles (red, blue and green) correspond to $j=0,1,2$.

The size of the gap is roughly proportional to the matrix element of $U$ between the two states on the crossing line.
The graphene's eigenstates are written in the $(A,B)$ spinor representation as
\begin{equation}
|\Vec{k}, s \rangle 
= \frac{1}{\sqrt{2}} 
\begin{pmatrix}
1 \\
-s e^{i \theta(\Vec{k})}
\end{pmatrix},
\end{equation}
where $s=\pm$ represent the conduction and valence bands, respectively,
and $\theta(\Vec{k}) = {\rm arctan} (k_y/k_x)$.
Now, the matrix element of $U$ from graphene 1 to graphene 2 is
\begin{align}
\langle \Vec{k}+\Delta\Vec{K}_j, + | U |\Vec{k}, - \rangle \approx i u \sin \Big[\theta(\Vec{k}) - \frac{2\pi j}{3}\Big],
\label{eq_mat_approx}
\end{align}
where $|\Delta\Vec{K}_j| \ll |\Vec{k}|$ is assumed.
In Fig.\ \ref{fig_FS}(b), the thickness of the crossing lines represents the amplitude of the interlayer matrix element
on the Dirac cones of layer 1 and 2, respectively.
We see  the matrix element vanishes near $E=0$, and this is the reason why 
the two major gaps open above and below $E=0$.
In a small $u$ limit, the number of states (per area) sandwiched by the two gaps is given by
$2 n_W$ where
\begin{equation}
n_W = g_vg_s \frac{\Delta}{4\pi \hbar v} \Delta K,
\end{equation}
where $g_v=g_s=2$ are the spin and valley degeneracies, and $\Delta K = |\Delta\Vec{K}_j|=4\pi\theta/(3a)$.
The $n_W$ characterizes the typical carrier density to reach the energy window of the 1D eigenmodes. 
For $\theta = 0.3^\circ$ and $\Delta =200$ meV, for instance,
we have $n_W = 1.08\times 10^{12}$cm$^{-2}$.

The 1D eigenmodes remaining inside the energy window can be explained by the reconstruction of the Fermi surface.
Figures \ref{fig_FS}(c) and (d) illustrate the Fermi surfaces before introducing $u$ at $E_F=20$ meV,
which is slightly below the maximum energy of the crossing rings.
In panel (c), the central dashed circle represents the hole band of layer 1, 
and the three solid circles are the electron band of layer 2 with three momentum shifts $\Delta \Vec{K}_j \, (j=0,1,2)$.
In panel (d), the Fermi surface of layer 1 is centered instead.
The hybridized Fermi surfaces after the infinitesimal anti-crossing are shown in Fig.\ \ref{fig_FS}(e).
We therefore have three open Fermi surfaces that are 120 degrees apart as well as three closed pockets.
By increasing $u$, the closed pockets are gapped out due to a good nesting between the electron and hole parts.
On the other hand, the open Fermi surfaces remain ungapped, 
which explains the origin of the 1D eigenmodes filling the gap.
We also see that the open Fermi surface mainly consists of the layer 1 component (solid line)
which is consistent with the fact that the wave function has larger amplitude on layer 1 in Fig.\ \ref{fig_wave}. 

\begin{figure}
        \center{
        \includegraphics[width=0.9\hsize]{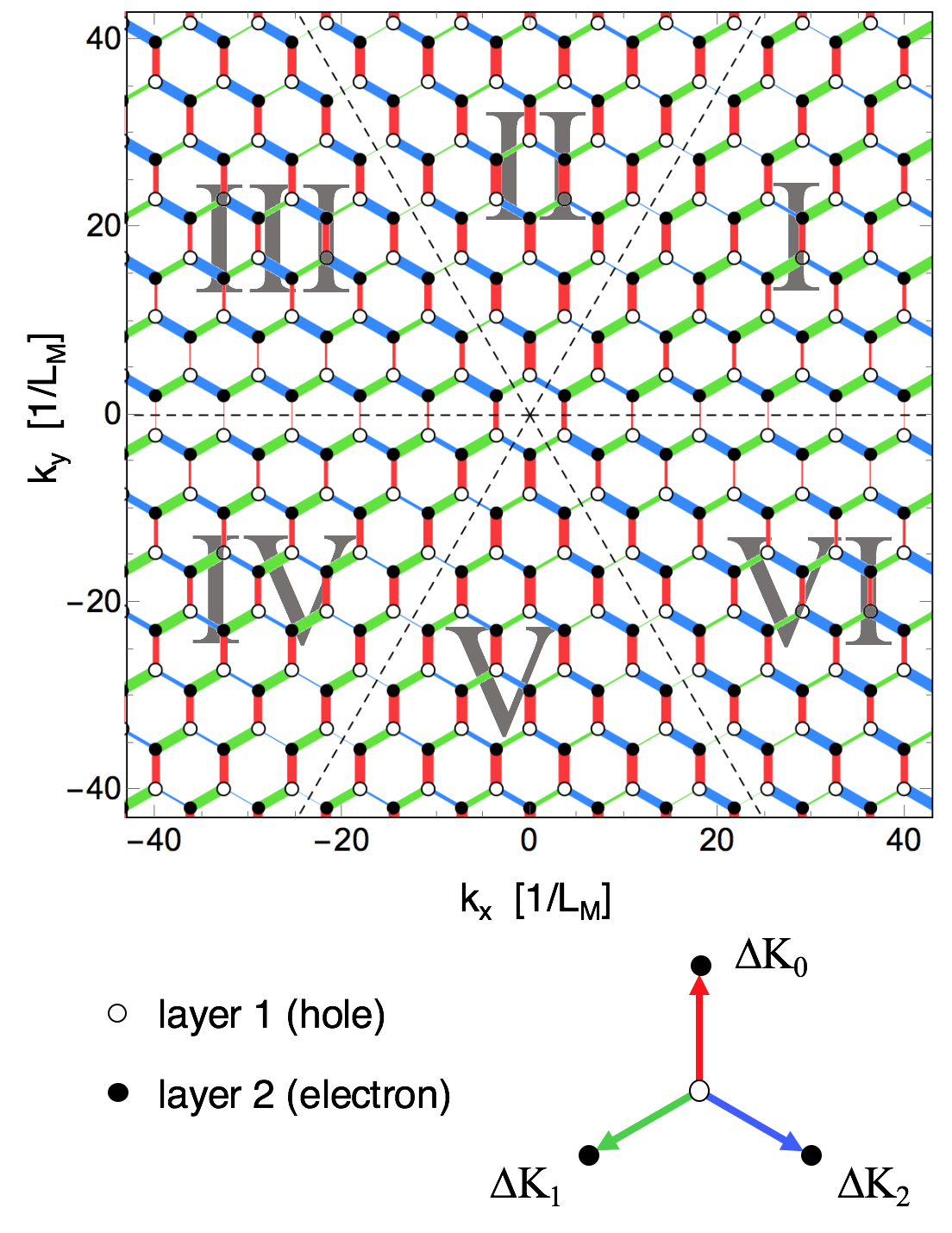}
        }
        \caption{
Map of the interlayer matrix element,
where open circles represent the hole states at $\Vec{k} ^{(1)}_{m_1,m_2}$ of single-layer graphene 1,
and filled circles the electron states at $\Vec{k} ^{(2)}_{m_1,m_2}$ of graphene 2, 
where $\Vec{k} =0$ is chosen in Eq.\ (\ref{eq_k_vectors}).
The bond thickness is proportional to the matrix element of $U$ between the two states.
} 
        \label{fig_network}
\end{figure}

In this picture, we only consider the band crossing of the first order in $u$,
while we actually have high-order hybridization at other crossing points.
It is somewhat surprising that the 1D eigenmodes in three directions are not hybridized and remain independent
even in a large $u$ beyond the pertubational regime. 
This is understood by the $k$-space map of the interlayer matrix element in Fig.\ \ref{fig_network},
where open circles represent the graphene 1's hole states at $\Vec{k} ^{(1)}_{m_1,m_2}$,
filled circles the graphene 2's electron states at $\Vec{k} ^{(2)}_{m_1,m_2}$ ,
and the bond thickness is proportional to the matrix element of $U$ between the two states. We can show that the 1D eigenmodes in the positive energy window are contributed 
by graphene's states only in the regions I, III and V,
and those in the negative energy window are 
by the regions II, IV and VI.
This is consistent with the observation that the open Fermi surface in Fig.\ \ref{fig_FS}(e) consists of
the graphene's Fermi surface portions in the same regions.
We see that the matrix element nearly vanishes on the boundary of different regions (dashed lines)
according to Eq.\ (\ref{eq_mat_approx}),
except for the $k$-points near the origin which do not contribute to the low-energy states.
As a result, the six regions I, II, $\cdots$ VI are nearly decoupled 
and that is why the 1D eigenmodes running in the different directions remain independent in increasing $u$.

The perfect 1D eigenmodes in the biased TBG
is analogous to those in zigzag graphene nanoribbons \cite{wakabayashi2007perfectly}.
In a doped zigzag nanoribbon, it is known that
each valley has different numbers of left-moving modes and right-moving modes
at the Fermi energy;
$n$ right modes and $n+1$ left modes at one valley, 
while $n+1$ right modes and $n$ left modes at the other valley.
The excess traveling mode in each valley remains as a perfectly conducting channel
even in the disordered system, as long as the impurities are long-ranged and do not mix the different valleys.
The perfect 1D eigenmodes in biased TBG can be viewed as a 2D version of this, in that
each single sector of I, III and V  has different numbers of out-going modes and in-coming modes (with respect to the graphene's Fermi circle),
as is clear from different arc lengths of the electron and hole Fermi surfaces in Fig.\ \ref{fig_FS}(e),
and that different sectors are not hybridized by the interlayer coupling $u$.
Therefore the excess modes originate from the electron Fermi surface and they remain as traveling modes
in the presence of $u$.

\section{Discussion}

The disorder effect on these 1D eigenmodes is an important problem
when considering the electronic transport.
As shown in Fig.\ \ref{fig_schem_domain}(b),
each 1D mode is composed of straight parts on the AB-BA boundary
and corner angles on AA spots.
A hybridization of different 1D eigenmodes takes place
only by a local mixing of the mode 1 and 2 on the AB-BA boundaries,
or an irregular reflection at the AA corners.
In real TBGs, the moir\'{e} structure exhibits a distorted triangular pattern 
with shifted AA spots and extended / shortened AB-BA boundaries \cite{brown2012twinning,lin2013ac,alden2013strain},
However, we expect that such a moir\'{e}-scale distortion would not 
cause a strong mixing of different 1D eigenmodes,
because the local atomic structures of AA and AB-BA boundary are not modified very much \cite{koshino2019moire},
such that the hubs and the links in the triangular lattice work in the same way as in the non-distorted system.  
Major scatterings should be mainly caused by short-ranged disorder
smaller than the local structures of AB-BA boundary and AA spots (which is about a few nm).
The detailed study on the disorder scattering will be left for future works.

When the scattering can be neglected and the Fermi energy is in the energy window, 
the electronic transport must be dominated by the ballistic transport through the 1D eigenmodes. It is also expected that we do not have the Aharanov-Bohm (AB) oscillation in magnetic fields,
because the 1D eigenmodes do not enclose triangular domains, so do not cause any interferences.
Recently, transport measurements have been performed on small-angle TBGs under interlayer bias,
and a significant AB oscillation was observed \cite{rickhaus2018transport,yoo2019atomic,giant2019oscillations}.
We expect that magnetic oscillations take place when the perfect 1D eigenmodes are not well formed,
e.g., because the bias is not enough or the Fermi energy is not in the corresponding region.
To have the perfect 1D eigenmodes, it is required that the energy window $(|E_F|\lsim u)$ is dominated by the hole band of a single layer and the electron band of the other layer,
and this gives a condition $\Delta/2 \gsim u$ i.e.,  $\Delta \gsim 200$ meV.

\section{Conclusion}
We used the continuum-model and tight-binding Hamiltonians to show that TBGs with an applied bias exhibit perfect 1D eigenmodes in well-defined energy windows on either side of zero energy. We found that these states never hybridise and they propagate independently in three different directions along the domain walls separating AB and BA regions. In the presence of arbitrary lattice deformations, we show that the wave functions become even more localized on the domain boundaries. 
The formation of the well-defined energy windows and the origin of these states is explained by the two-band model
consisting of the intersecting electron and hole bands of single layer graphene,
where the 1D eigenmodes correspond to the emergent open Fermi surfaces formed by the moir\'{e} interlayer hybridization.

%We leave the detailed study of the nearly flat bands forming  in the upper part of the energy window for future work.

\section{Acknowledgments}

BT acknowledges financial support from the Japan Society for the Promotion of Science as a JSPS International Research Fellow (Summer Programme 2019), EPSRC Doctoral Training Centre Graphene NOWNANO EP/L01548X/1 and the Lloyd's Register Foundation Nanotechnology grant.
MK acknowledges the financial support of JSPS KAKENHI Grant Number JP17K05496. 

\bibliography{1d_channel}

\end{document}